\def\paperID{0019}
\def\confName{CVPR}
\def\confYear{2025}

\def\paperTitle{COP-GEN-Beta: Unified Generative Modelling of COPernicus Imagery Thumbnails}

\def\authorBlock{
    Miguel Espinosa$^{1,2}$\thanks{First and corresponding author (miguel.espinosa@ed.ac.uk)\\Work done during an internship at ESA $\Phi$-Lab.} \quad
    Valerio Marsocci$^2$ \quad
    Yuru Jia$^3$ \quad
    Elliot J. Crowley$^1$ \quad
    Mikolaj Czerkawski$^{2,4}$ \\
    \\
    $^1$University of Edinburgh \quad
    $^2$European Space Agency (ESA) \quad
    $^3$KU Leuven \quad
    $^4$Asterisk Labs
}

\newif\ifreview 
\newif\ifarxiv \newcommand{\arxiv}{\arxivtrue}
\newif\ifcamera 
\newif\ifrebuttal 

\arxiv % \review OR \arxiv OR \cameraready

\pdfoutput=1
\documentclass[10pt,twocolumn,letterpaper]{article}
\ifreview \usepackage[review]{cvpr} \fi
\ifarxiv \usepackage[pagenumbers]{cvpr} \fi
\ifrebuttal \usepackage[rebuttal]{cvpr} \fi
\ifcamera \usepackage{cvpr} \fi

\usepackage{graphicx}	
\usepackage{amsmath}	
\usepackage{amssymb}	
\usepackage{booktabs}
\usepackage{times}
\usepackage{microtype}
\usepackage{epsfig}
\usepackage{caption}
\usepackage{float}
\usepackage{placeins}
\usepackage{color, colortbl}
\usepackage{stfloats}
\usepackage{enumitem}
\usepackage{tabularx}
\usepackage{xstring}
\usepackage{multicol}
\usepackage{multirow}
\usepackage{xspace}
\usepackage{url}
\usepackage{subcaption}
\usepackage{xcolor}
\usepackage[hang,flushmargin]{footmisc}

\definecolor{mygreen}{RGB}{129,222,118}
\definecolor{myblue}{RGB}{108,173,223}
\definecolor{myred}{RGB}{204,0,0}

\usepackage{xspace}
\newcommand{\ourmodel}{COP-GEN-Beta\xspace}

\ifcamera \usepackage[accsupp]{axessibility} \fi

% ...

\ifarxiv  \fi

\newcommand{\R}[1]{{%
    \textbf{%
        \ifstrequal{#1}{1}{\textcolor{red}{R#1}}{%
        \ifstrequal{#1}{2}{\textcolor{blue}{R#1}}{%
        \ifstrequal{#1}{3}{\textcolor{magenta}{R#1}}{%
        \ifstrequal{#1}{4}{\textcolor{teal}{R#1}}{%
                           \textcolor{cyan}{R#1}%
        }}}}%
    }%
}}

\usepackage{xr-hyper}

\makeatletter
\newcommand*{\addFileDependency}[1]{
  \typeout{(#1)}
  \@addtofilelist{#1}
  \IfFileExists{#1}{}{\typeout{No file #1.}}
}

\makeatother
\newcommand*{\myexternaldocument}[1]{
    \externaldocument{#1}
    \addFileDependency{#1.tex}
    \addFileDependency{#1.aux}
}

\definecolor{cvprblue}{rgb}{0.21,0.49,0.74}
\usepackage[pagebackref,breaklinks,colorlinks,allcolors=cvprblue]{hyperref}
\usepackage[capitalize]{cleveref}
\crefname{section}{Sec.}{Secs.}
\crefname{table}{Table}{Tables}
\crefname{figure}{Fig.}{Figs.}

\ifarxiv \crefname{appendix}{App.}{Apps.}
\else \crefname{appendix}{Suppl.}{Suppls.} \fi

\unless\ifarxiv \myexternaldocument{_supplementary} \fi

\begin{document}
%% TITLE
\title{\paperTitle}
% Or COP-G
% Or COP-T (thumbnail)
% And the full model COP-GEN

\author{\authorBlock}
\maketitle

\begin{abstract}

    In remote sensing, multi-modal data from various sensors capturing the same scene offers rich opportunities, but learning a unified representation across these modalities remains a significant challenge. Traditional methods have often been limited to single or dual-modality approaches. In this paper, we introduce \ourmodel, a generative diffusion model trained on optical, radar, and elevation data from the Major TOM dataset. What sets \ourmodel apart is its ability to map any subset of modalities to any other, enabling zero-shot modality translation after training. This is achieved through a sequence-based diffusion transformer, where each modality is controlled by its own timestep embedding. We extensively evaluate \ourmodel on thumbnail images from the Major TOM dataset, demonstrating its effectiveness in generating high-quality samples. Qualitative and quantitative evaluations validate the model’s performance, highlighting its potential as a powerful pre-trained model for future remote sensing tasks.
    
    Project page:~\url{http://miquel-espinosa.github.io/cop-gen-beta}
\end{abstract}

{\begin{figure}[ht]
    \centering
    \includegraphics[width=\linewidth]{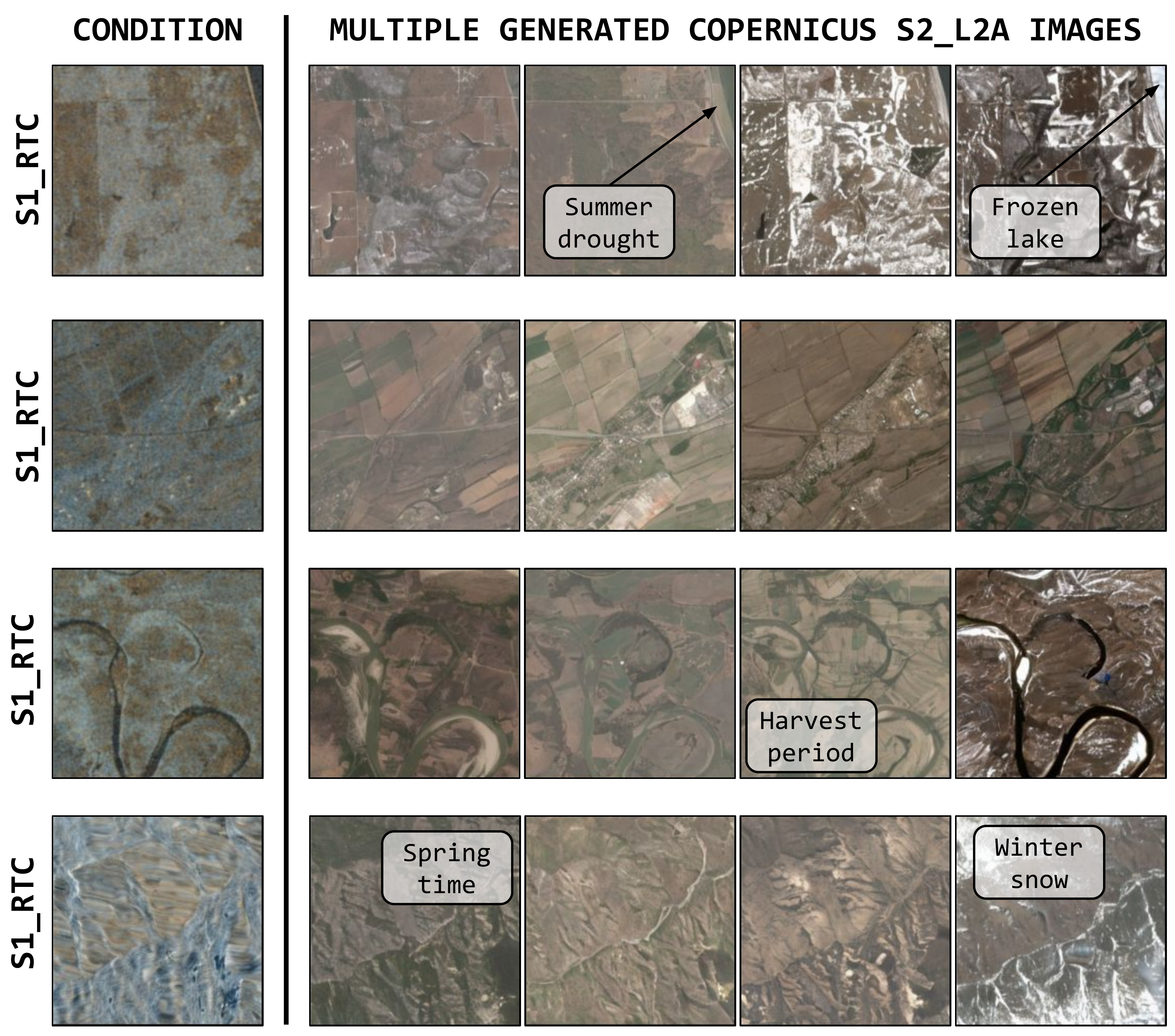}
    \caption{By training on dense, global coverage \ourmodel has captured a wide and diverse data distribution of the supported modalities. It is possible to observe emergent effects such as seasonality when sampling multiple images conditioned on the same S1RTC sample, despite having trained on only one temporal sample for each location in the world (since Major TOM does not provide multi-temporal data). \ourmodel is capable of synthetising new locations that do not exist, but also it can reimagine existing locations in conditions that were never observed. Best viewed when zoomed in.}
    \label{fig:emergent-seasonality}
\end{figure}}

{% Use figure* for multi-column figure
\begin{figure*}[ht]
    \centering
    \includegraphics[width=\linewidth]{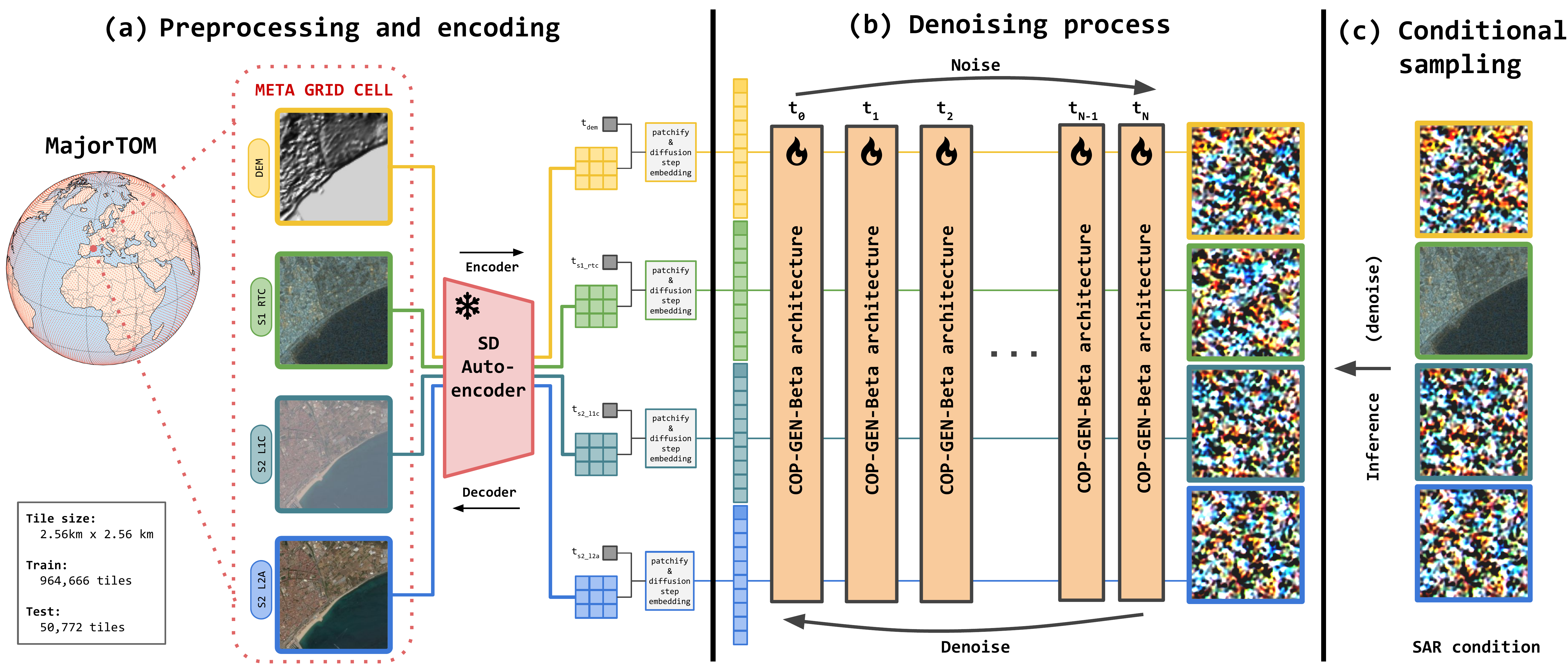}
    \caption{\ourmodel is the first generative model trained on the joint distribution of Sentinel-2 (both L1C and L2A), Sentinel-1 RTC, and Copernicus GLO-30 DEM data. This is done through (a) sampling a global and dense dataset of these modalities from Major TOM, encoding all images with a pretrained StableDiffusion autoencoder, and (b) training a sequence-based denoising diffusion model using a transformer backbone, where each modality is supplied with its designated timestep. This approach makes it possible to (c) generate all modalities based on any subset thereof that is available.}
    \label{fig:teaser-architecture}
\end{figure*}
}

\section{Introduction}
\label{sec:intro}

    The Copernicus programme is one of the largest and most extensively used sources of open Earth observation data, designed for global monitoring of the planet and enabling a wide array of applications~\cite{aschbacher2017esa}. It is one of the most prominent examples of large-scale multi-modal data (including optical, radar, and elevation modalities), delivering petabytes of openly accessible data every year, free to use by anyone~\cite{copernicus2022}. This amount of information about the state of Earth is naturally challenging to process and interpret without automation, and consequently, methodologies leveraging the recent developments in the field of artificial intelligence and computer vision have been playing a substantial role~\cite{apicella2022copernicus}.
    
    A prominent challenge in the field of Earth observation is the alignment of multiple sensor modalities, ensuring that the information about a scene from different sources captured at different times can be integrated into the same analysis pipeline~\cite{miller2024deep}.
    Improving an interface for diverse modalities can support many areas that are currently facing this challenge~\cite{zhang2010multi, li2022deep}, such as climate monitoring~\cite{marsocci2024pangaea},~disaster response \cite{kumar2023state}, or land-use classification~\cite{van2018spacenet}. Furthermore, there is a growing need for general-purpose pre-trained models that can serve as a powerful source of prior for the data modalities of interest, similarly to the way pre-trained generalist models have been a fundamental tool in the mainstream domain of deep learning~\cite{Bommasani2021FoundationModels}.
    
    Despite the abundance of open data in this area, general-purpose generative models for Earth observation remain limited in number and scope. In domains such as natural language processing and computer vision, pre-trained generative models have become essential tools. In contrast, the Earth observation community has yet to fully leverage the power of generative modeling to capture the complex statistical relationships between multiple remote sensing modalities and the majority of previous works on the topic consider a single modality at a time (most often optical)~\cite{liu2024diffusion,khanna2024diffusionsat}.

    This work aims to explore the potential of highly multimodal generative models in the field of Earth observation. The goal is to develop a solution that enables flexible translation between any combination of supported data modalities. Achieving this requires overcoming several key challenges, the most important one being the identification, implementation, and training of a suitable architecture on a diverse and comprehensive dataset.

    This work presents a solution to this challenge: \ourmodel, a denoising diffusion model based on a transformer architecture, which allows for multiple modalities to be injested as a sequence of latent tokens, combined with individual diffusion timestep embeddings that provide full control of the generative process. 
    
    By adjusting the diffusion time for each modality, users can determine which modalities are generated and even control how much of the diffusion chain is traversed during generation. This flexibility allows for conditional generation based on any subset of available modalities---or, in the absence of conditions, fully unconditional generation.
    \ourmodel is trained on a globally distributed and densely sampled dataset from Copernicus, using the Major TOM dataset thumbnails~\cite{francis2024major}. The diversity of the training set enables the model to learn robust and globally applicable representations, as illustrated in Figure~\ref{fig:emergent-seasonality}.

    Both quantitative and qualitative evaluations demonstrate that \ourmodel has learned a powerful representation of four distinct data modalities, highlighting the potential of transformer-based diffusion approaches for multimodal generative modelling in Earth observation. These findings suggest a promising direction for future models, extending beyond the exploratory domain of thumbnail-based generation.
    
    The main contributions of this work are as follows: 
    \begin{itemize}
        \item \ourmodel is introduced, a generalist generative model for Major TOM thumbnails, capable of modeling the joint distribution of four modalities from the Copernicus programme.
        \item \ourmodel gives full flexiblity of generation, capable of translating between any subset of supported domains and establishing a new state-of-the-art in terms of versatile generative models for Earth observation data.
        \item The unique capabilities of \ourmodel are demonstrated through quantitative and qualitative evaluation.
    \end{itemize}
    
    This work represents a first step towards powerful and truly generalist generative models for Copernicus data, with promising results that motive further development in this direction.

\section{Related Work}
\label{sec:related}

    Given the multi-modal generative nature of \ourmodel, in Section \ref{sec:gen-eo} an overview of generative models for Earth observation (EO) is provided, while Section \ref{sec:gen-mm} gives a summary of multi-modal generative models.

    \subsection{Generative Models in Earth Observation}
    \label{sec:gen-eo}
    
    The concept of training generative models on EO data is not new and has been explored for several years now. The early works on the topic employed primarily generative adversarial networks (GANs) \cite{wang2023review, li2021deep, jiang2019edge} and normalizing flows \cite{wu2023conditional, enescu2024conditional, li2023disasternet}.

    More recently, several diffusion models have been proposed for satellite imagery~\cite{Sanguini2023, liu2024diffusion, le2024detecting,khanna2024diffusionsat,espinosa_2023_8_mapsat,metaearth,zheng2024changen2,tang2024crs}. MetaEarth~\cite{metaearth} employs a resolution-guided approach to generate images of arbitrary sizes, while DiffusionSat~\cite{khanna2024diffusionsat} creates satellite imagery based on semantic text and metadata. CRS-Diff~\cite{tang2024crs} takes it further by improving the controllability of scene generation through the integration of multiple conditioning inputs. In addition to generative tasks, diffusion models have also been successfully applied to various image enhancement tasks, such as denoising~\cite{he2023tdiffde,pang2024hir}, cloud removal~\cite{wang2024idf,zou2024diffcr}, super-resolution~\cite{wang2025semantic,dong2024building}, or terrain generation~\cite{bornepons2025mesa} demonstrating the versatility of generative modelling in remote sensing (RS) applications.

    General-purpose visual generative models can also be used for processing Earth observation data, as satellite imagery does appear in large-scale datasets derived from the internet~\cite{czerkawski2023laion} that are often used as training data. However, the representation of Earth observation data on the internet is heavily biased~\cite{czerkawski2023laion} and even if these models can sometimes provide a useful source of visual priors~\cite{Czerkawski_2024}, their performance on satellite imagery often falls short compared to domain-specific models. Notably, satellite datasets such as EuroSAT have proven to be particularly challenging for early versions of general-purpose models like CLIP \cite{radford2021learning}.

    \subsection{Multi-modal models} 
    \label{sec:gen-mm}

    Multi-modal learning is a widely studied topic in computer vision \cite{han2024multimodal}, with research often focusing on combinations of optical and depth data \cite{zhao2021adaptive, fooladgar2019multi} or vision and language models \cite{yang2024mma, wu2025visionllm}. While these approaches have been effective for specific tasks \cite{zhang2024vision, ye2024cat}, they typically model only a single-directional mapping between modalities \cite{ren2024timechat, han2024multimodal} rather than learning a joint distribution that allows sampling from marginal distributions. This limitation is often observed in multi-modal generative models, where, most of the time, the generation is focused on one modality \cite{Ruiz_2023_CVPR}, sometimes using multi-modal conditioning \cite{Huang_2023_CVPR, Hu_2024_CVPR}. Some research works have investigated generation of modality pairs, like text-vision \cite{gao2023textpainter} or video-audio \cite{Ruan_2023_CVPR}. Few approaches propose a unified framework for joint multi-modal generation. Among them, UniDiffuser \cite{pmlr-v202-bao23a} particularly stands out. UniDiffuser -- still tested just on vision-language -- can learn joint distributions, using a Transformer to diffuse them \cite{Bao_2023_CVPR}. 

    Based on this, \ourmodel, inspired by \cite{pmlr-v202-bao23a}, learns through diffusion a joint distribution of Earth observation data. Unlike previous generative models in this domain, \ourmodel can be used in a zero-shot manner to translate between any subset of learned modalities without requiring explicitly paired training data. As satellite data volume and sensor diversity continue to grow, introducing joint diffusion models to Earth observation offers new possibilities for sensor fusion, data harmonization, and improved downstream analysis.

\section{Background}

\paragraph{Diffusion models.} Diffusion models \cite{ho2020denoising} generate data by learning to reverse a noise corruption process. The forward process gradually adds Gaussian noise to data in a Markov chain:  
\[
q(x_t | x_{t-1}) = \mathcal{N}(\sqrt{\alpha_t} x_{t-1}, \beta_t I),
\]
where \( \alpha_t = 1-\beta_t\) and \( \beta_t \) define the noise schedule. To generate new samples, the model learns an approximate reverse process, modeled as a Gaussian distribution \( p(x_{t-1} | x_t) \) with mean \( \mu_t(x_t) \) and variance \( \sigma_t^2 \). The optimal mean estimator \cite{pmlr-v202-bao23a} is given by:
\[
\mu_t^*(x_t) = \frac{1}{\sqrt{\alpha_t}} \left( x_t - \frac{\beta_t}{\sqrt{1 - \alpha_t}} \mathbb{E}[\epsilon | x_t] \right),
\]
where \( \epsilon \) represents the added noise. Learning diffusion models thus reduces to training a neural network \( \epsilon_\theta(x_t, t) \) to predict \( \epsilon \) by minimizing the objective:
\[
\min_{\theta} \mathbb{E}_{t, x_0, \epsilon} \|\epsilon - \epsilon_\theta(x_t, t)\|_2^2.
\]
For conditional generation, auxiliary information \( c \) is incorporated into the model, extending the noise prediction task to:
\[
\min_{\theta} \mathbb{E}_{t, x_0, c, \epsilon} \|\epsilon - \epsilon_\theta(x_t, t, c)\|_2^2.
\]

\paragraph{Conditional generation across modalities.}  
Given data from \( M \) modalities, \( (\mathbf{x}_0^{(1)}, \dots, \mathbf{x}_0^{(M)}) \sim q(\mathbf{x}_0^{(1)}, \dots, \mathbf{x}_0^{(M)}) \), we aim to model the conditional distribution of one modality given the others. The reverse process follows a Gaussian model, where the mean is estimated as:  
\[
\boldsymbol{\mu}_t^*(\mathbf{x}_t^{(i)} | \mathbf{x}_t^{(j)}) = \frac{1}{\sqrt{\alpha_t}} \left( \mathbf{x}_t^{(i)} - \frac{\beta_t}{\sqrt{1 - \alpha_t}} \mathbb{E}[\boldsymbol{\epsilon}^{(i)} | \mathbf{x}_t^{(i)}, \mathbf{x}_t^{(j)}] \right).
\]
A noise prediction network \( \boldsymbol{\epsilon}_\theta \) conditioned on cross-modal information is trained to minimize:
\[
\min_{\theta} \mathbb{E}_{t, \mathbf{x}_0, \boldsymbol{\epsilon}} \|\boldsymbol{\epsilon}^{(i)} - \boldsymbol{\epsilon}_\theta (\mathbf{x}_t^{(i)}, \mathbf{x}_t^{(j)}, t)\|_2^2.
\]
This enables diffusion models to learn structured cross-modal representations for generative tasks.

\section{COP-GEN-Beta: Generative Modelling of Copernicus Imagery}
\label{sec:method}

We introduce \ourmodel, a novel diffusion model designed to handle multiple remote sensing modalities. Specifically, \ourmodel operates on four key EO modalities: Digital Elevation Model (DEM), Sentinel-1 Radar Terrain Corrected (S1 RTC), Sentinel-2 Level 1C (S2 L1C), and Sentinel-2 Level 2A (S2 L2A). Unlike previous approaches, which require separate models for per modality, \ourmodel learns joint, conditional, and marginal distributions within a unified framework. The model architecture is illustrated in Figure \ref{fig:teaser-architecture}.

\subsection{Dataset details}

The dataset is a global collection, covering the entire Earth's surface, and is extracted from MajorTOM~\cite{francis2024major}. It consists of four modalities: Sentinel-2 Level 2A (S2L2A), Sentinel-2 Level 1C (S2L1C), Sentinel-1 (S1), and Digital Elevation Model (DEM) (from the Core-DEM expansion dataset~\cite{bornepons2025mesa}). Both the S2L1C and S2L2A data share the same characteristics: they are RGB composites (B04, B03, B02) at a 10m resolution, saved as 1068x1068 PNG files, with a 256x256 pixel center-cropped patch extracted for each thumbnail. S2L2A data includes atmospheric correction, whereas S2L1C data does not. The S1 data consists of rescaled false-color images at a 10m resolution, saved as 1068x1068 PNGs, with RGB channels defined as R:VV, G:VV+VH, and B:VH. Normalization applied to the S1 data means that values are relative to each individual thumbnail, so they cannot be compared across different thumbnails. The DEM modality consists of a single greyscale channel with a compressed hillshade visualization at a 30m resolution, saved as 356x356 images, and is rescaled to align with the S1 and S2 datasets. Sourced from the Copernicus Digital Surface Model (DSM), the DEM represents the Earth's surface, including buildings, infrastructure, and vegetation.
Splitting the dataset yields train and test subsets of respectively 964,666 and 50,772 tiles.

\subsection{Preprocessing and Latent Encoding}

    Given a cropped input image $\mathbf{x}^{(i)}$ of shape $(3, 256, 256)$ from modality $i$, where $i \in {1, \dots, M}$ and $M=4$, we encode it using a pretrained frozen encoder $\mathcal{E}$ (specifically, from Stable Diffusion\footnote{\url{https://github.com/CompVis/stable-diffusion}} \cite{Rombach_2022_CVPR}). Sharing the same frozen encoder-decoder among all input modalities allows the model to maintain a common latent space across different sensor types, enforcing a unified representation. 
    
    For each image $\mathbf{x}^{(i)}$, the encoder $\mathcal{E}$ produces a latent distribution with mean $\boldsymbol{\mu}^{(i)}$ and variance $\boldsymbol{\sigma}^{(i)}$.

    We sample $\mathbf{z}^{(i)} \sim \mathcal{N}(\boldsymbol{\mu}^{(i)}, \boldsymbol{\sigma}^{(i)})$, where $\mathbf{z}^{(i)} \in \mathbb{R}^{4 \times 32 \times 32}$. These latents are then used as inputs to the diffusion process.
\subsection{Diffusion process}
    
    \paragraph{Timestamps and Modality Interaction.} A key aspect of \ourmodel is its independent timestamp encoding for each modality. During training, each modality $M$ is assigned separate diffusion timesteps $t^{(i)}$, encoded through a modality-specific time embedding module $\phi^{(i)}$. Having timestamp independence enables the model, given sufficient training, to differentiate between modalities while learning their distinct characteristics within a shared backbone. Additionally, it allows to sample different distributions at inference time by adjusting the timesteps of individual modalities separately (See "Sampling Modes"). Meanwhile, the shared backbone promotes cross-modal information exchange (attention mechanism), enhancing both generative and discriminative downstream performance.

% \subsection{\ourmodel Architecture}

\paragraph{Training.} During training, given $M$ latents $\mathbf{z}^{(i)}$, we apply a modality-specific patch embedding layer $P_i$ to transform them into token embeddings:
\begin{equation}
\mathbf{h}^{(i)} = P_i(\mathbf{z}^{(i)}), \quad \mathbf{h}^{(i)} \in \mathbb{R}^{N \times d}, \quad i \in 1,...,M
\end{equation}
where $N$ is the number of tokens per modality, and $d$ is the embedding dimension. Similarly, timestamps $t^{(i)}$ are processed through $\phi^{(i)}$ (a small MLP) to obtain modality-specific time embeddings:
$$\mathbf{e}_i = \phi^{(i)}(t^{(i)})$$

All token embeddings from all modalities are then concatenated into a single sequence, constituting the input to the transformer blocks:
$$\mathcal{X} = [\mathbf{e}_1, \mathbf{e}_2, ..., \mathbf{e}_M, \mathbf{h}_1, \mathbf{h}_2, ..., \mathbf{h}_M]$$

Each transformer block contains: (1) Multi-head self-attention (MHSA) to capture global dependencies across modalities, (2) Modality-specific dropout and layer normalization, (3) Shallow-to-deep skip connections to preserve spatial structure, critical for generative models, (4) Feed-forward MLP layers with GELU activation. The output tokens are split to recover the individual modality representations.

\paragraph{Unified Noise Prediction.} Given the noisy latents $\mathbf{z}_{t}^{(i)}$ at time step $t^{(i)}$, our transformer-based backbone $F_{\theta}$ predicts the corresponding noise components $\hat{\boldsymbol{\epsilon}}^{(i)}$ for each modality.

\begin{equation}
\hat{\mathcal{X}} = F_{\theta}([\mathbf{e}_1, \mathbf{e}_2, ..., \mathbf{e}_M, \mathbf{h}_1, \mathbf{h}_2, ..., \mathbf{h}_M])
\end{equation}

From the predicted token sequence $\hat{\mathcal{X}}$, we extract the predicted latent noise components for each modality, $\hat{\boldsymbol{\epsilon}}^{(i)}$.

The training objective remains a DDPM-like regression loss:

\begin{equation}
    \mathcal{L} = \mathbb{E}_{\mathbf{z}_0, t, \boldsymbol{\epsilon}} \left[ \sum_{i=1}^{M} \left\| \boldsymbol{\epsilon}^{(i)} - \hat{\boldsymbol{\epsilon}}^{(i)} \right\|^2 \right]
\end{equation}

where $\boldsymbol{\epsilon}^{(i)}$ represents the true noise added to $\mathbf{z}^{(i)}$ during training.

\paragraph{Inference.} At inference time, the diffusion steps are applied iteratively across $t$, refining the noisy latents until $t=0$:
\begin{equation}
\hat{\mathbf{z}}_{t-1}^{(i)} = \text{Transformer}(\mathbf{z}_t^{(i)}).
\end{equation}

Finally, the decoder $\mathcal{D}$ reconstructs the original image from the denoised latent:
\begin{equation}
\hat{\mathbf{x}}^{(i)} = \mathcal{D}(\hat{\mathbf{z}}_0^{(i)}),
\end{equation}
where $\hat{\mathbf{z}}_0^{(i)}$ is the estimated clean latent at $t=0$.

\subsection{Sampling modes}

    \subsubsection{Joint unconditional generation}
    In joint generation mode, we sample all modalities simultaneously. This corresponds to sampling from the joint distribution $q(\mathbf{z}_1, \mathbf{z}_2, ..., \mathbf{z}_M)$. The process begins with randomly sampled latents for each modality, and the diffusion model iteratively denoises all modalities simultaneously, with each modality having the same timestep schedule.
    This setting is particularly useful for generating synthetic multi-modal paired datasets, where alignment between modalities is crucial.

    \subsubsection{Conditional generation}
    
    In conditional generation mode, we generate a subset of modalities conditioned on the others. For example, given a Sentinel-2 L1C image, we can generate the corresponding DEM, Sentinel-1 RTC, and Sentinel-2 L2A images.
    To implement this, we partition the modalities into two non-overlapping sets: the conditioning set $C$ and the generation set $G$. For modalities in $C$, we keep their timesteps to $0$, effectively treating them as noiseless conditioning information. For modalities in $G$, we perform the standard diffusion process, starting from random noise and iteratively denoising.
    Formally, the conditional sampling process is:
    \begin{enumerate}
        \item Initialize random noise for modalities in $G$: $\mathbf{z}_T^{(i)}T \sim \mathcal{N}(0, I)$ for $i \in G$.
        \item Set modalities in $C$ to the conditioning encoded images: $\mathbf{z}^{(i)}_0 = \mathbf{z}^{(i)}$ for $i \in C$.
        \item For each timestep $t$ from $T$ to $1$:
        \begin{enumerate}
            \item Predict noise for modalities in $G$ using all modalities as input
            \item Update $\mathbf{z}_{t-1}^{(i)}$ for $i \in G$ using the predicted noise
            \item Keep $\mathbf{z}_0^{(i)}$ unchanged for $i \in C$
        \end{enumerate}
    \end{enumerate}
    This highly-flexible sampling scheme allows us to perform various downstream transformation tasks, such as:
    \begin{itemize}
        \addtolength{\itemindent}{0.1cm}
        \item Optical-to-radar translation (Sentinel-2 → Sentinel-1)
        \item Elevation estimation (Sentinel-1/2 → DEM)
        \item Atmospheric correction (Sentinel-2 L1C → Sentinel-2 L2A)
    \end{itemize}

{\begin{figure}[ht]
    \centering
    \includegraphics[width=\linewidth]{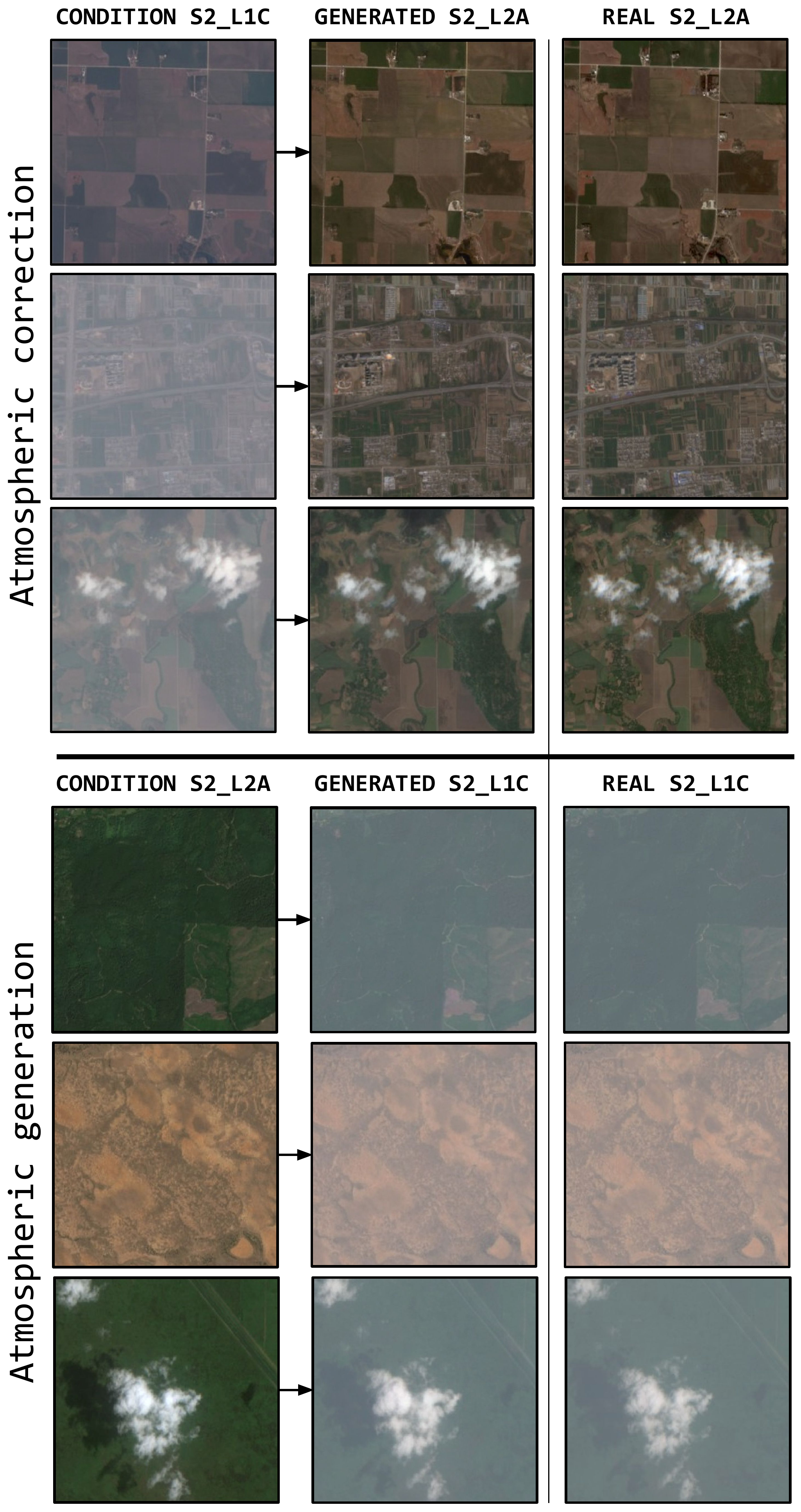}
    \caption{\ourmodel supports translation between processing levels of Sentinel-2, which emulates the official procesor for the L2A level (approximating Bottom-of-Atmosphere based on Top-of-Atmosphere input), but it can also do the opposite and approximate a possible L1C product from an observed L2A observation, which is equivalent to the synthesis of the atmospheric effect. Best viewed when zoomed in.}
    \label{fig:atmospheric-correction}
\end{figure}}

\subsection{Implementation Details}

\ourmodel is implemented as a Vision Transformer, closely following the design choices of \cite{pmlr-v202-bao23a}. We use a patch size of 2, an embedding dimension of 512, and a depth of 12 transformer layers (6 in-blocks and 6 out-blocks). For training, we use the Adam optimizer with a learning rate of 1e-4 and a cosine learning rate schedule. We employ Exponential Moving Average (EMA) of the model weights with a decay rate of 0.9999 to improve stability. The model is trained with an effective batch size of 2048 for $\sim$120,000 steps.
To reduce the computational cost and memory requirements, we pre-compute the input image latents using frozen pre-trained $E$, and train the diffusion model in this latent space. During inference, we use the DPM-Solver++ for efficient sampling with only 50 denoising steps.
Our implementation is built using PyTorch and is trained on 16 NVIDIA A100 (40GB) GPUs.

\subsection{Why COP-GEN-Beta? Advantages of a unified multi-modal generative model}

Existing diffusion models in the remote sensing landscape often focus on specific modality translations or require separate models for different data types. In contrast, \ourmodel introduces a unified, multimodal approach that enhances representation learning and generation capabilities.

First, it supports user-defined behavior, allowing the selection of known modalities while reconstructing missing ones. This eliminates the need for specialized models for specific translations (e.g. Sentinel-2 to Elevation) as long as the model supports the modalities of interest. This level of flexibility is particularly useful in applications where data availability is often inconsistent.

Second, by integrating a large number of related modalities, \ourmodel captures cross-modal relationships through a shared backbone, leveraging intermodal correlations to improve both data representation and generation. While similar strategies have been explored in representation learning \cite{fuller2023croma, nedungadi2024mmearth, Hong2024spectralgpt}, generative models have largely remained unimodal or narrowly focused on specific tasks. A model trained on joint multi-modal distributions provides a powerful prior for environmental interpretation from space.

Finally, the sequence-based nature of \ourmodel opens a door to easy introduction of new modalities by retraining the same model on sequences with additional input tokens. This way the model can integrate new data sources, with attention mechanisms allowing flexible cross-modal interactions without predefined constraints. The specific purpose of \ourmodel is to provide a useful set of pre-training weights for more complex versions of the model, taking advantage of a trade-off between low-cost thumbnail training (pre-trained autoencoder, 3 channels only) and the more expensive native support of original data representations.

{\begin{figure}[ht]
    \centering
    \includegraphics[width=\linewidth]{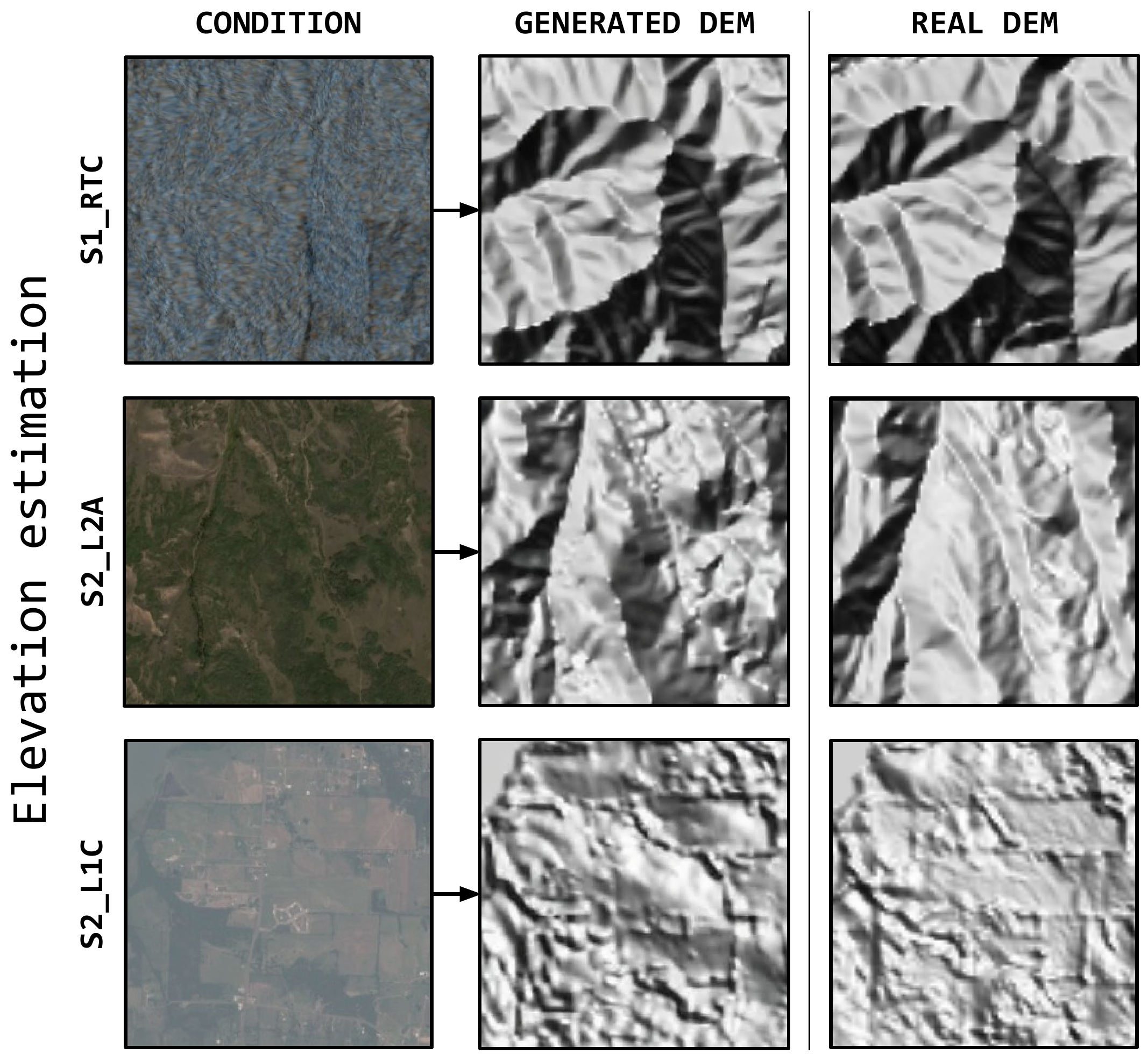}
    \caption{\ourmodel can map any modality to an elevation model estimate, which can be highly useful for dynamically changing terrains. Here, it is shown how any modality (Sentinel-1 or Sentinel-2) can be used to approximate the elevation, even radiometrically terrain-corrected radar product. Best viewed when zoomed in.}
    \label{fig:elevation-estimation}
\end{figure}}

\section{Evaluation}
\label{sec:experiments}

{
\begin{table*}[h]

    \centering
    \resizebox{\textwidth}{!}{
    \renewcommand{\arraystretch}{1.2} % Increase row height
    \setlength{\tabcolsep}{3pt} % Adjust column spacing

    \begin{tabular}{l c cccc cccc cccc cccc}
        \toprule
        \multicolumn{1}{c}{\multirow{2}{*}{\textbf{Model}}} & \multirow{2}{*}{\textbf{Condition}} & \multicolumn{4}{c}{\textbf{F-Score} $\uparrow$ (\%)} & \multicolumn{4}{c}{\textbf{Precision} $\uparrow$ (\%)} & \multicolumn{4}{c}{\textbf{Recall} $\uparrow$ (\%)} & \multicolumn{4}{c}{\textbf{FID} $\downarrow$} \\
        \cmidrule(lr){3-6} \cmidrule(lr){7-10} \cmidrule(lr){11-14} \cmidrule(lr){15-18}
        & & \textbf{DEM} & \textbf{S1RTC} & \textbf{S2L1C} & \textbf{S2L2A} & \textbf{DEM} & \textbf{S1RTC} & \textbf{S2L1C} & \textbf{S2L2A} & \textbf{DEM} & \textbf{S1RTC} & \textbf{S2L1C} & \textbf{S2L2A} & \textbf{DEM} & \textbf{S1RTC} & \textbf{S2L1C} & \textbf{S2L2A} \\
        \midrule
        % COP-GEN-real    &               &    &    &    &    & -        & -        & -        & -        & & & & & & & & \\
        % \midrule
        DiffusionSat \cite{khanna2024diffusionsat} & text prompt & -              & -              & 1.37           & 3.43           & -              & -              &  0.73          & 1.97           & -     &     - & 10.04 & 13.32 & -     & -     & 253.31 & 216.37 \\
        \ourmodel                                  & uncond.     & 42.25          & 24.22          & 36.71          & 39.54          & 73.25          & \textbf{49.81} & 73.65          & 71.72          & 29.69 & 16.00 & 24.45 & 27.29 & 90.36 & 52.21 &  53.66 &  48.49 \\
        \midrule                                                                                                                                                            
        \ourmodel                                  & DEM         & NA             & 30.14          & 42.71          & 46.64          & NA             & 45.35          & 68.04          & 67.77          &  NA   & 22.58 & 31.12 & 35.56 & NA    & 23.97 &  27.33 &  25.32 \\
        \ourmodel                                  & S1RTC       & 49.17          & NA             & 42.67          & 46.63          & \textbf{65.53} &  NA            & 68.96          & 69.38          & 39.35 & NA    & 30.89 & 35.12 & 49.27 & NA    &  29.55 &  27.18 \\
        \ourmodel                                  & S2L1C       & \textbf{51.89} & 32.57          & NA             & \textbf{83.51} & 64.25          & 47.22          &  NA            & \textbf{85.68} & 43.52 & 24.86 & NA    & \textbf{81.45} & 40.90 & \textbf{20.94} & NA     &   \textbf{7.71} \\
        \ourmodel                                  & S2L2A       & 51.41          & \textbf{33.13} & \textbf{82.78} & NA             & 62.71          & 46.75          & \textbf{85.50} &  NA            & \textbf{43.56} & \textbf{25.66} & \textbf{80.24} &  NA   & \textbf{36.70} & 20.98 &   \textbf{5.57} & NA     \\
        \bottomrule
    \end{tabular}
    }

    \caption{\textbf{Quantitative comparison of generative models on Copernicus data.} We compare our model (\ourmodel) under different conditioning settings against existing generative models. Metrics include F-Score, Precision, Recall (higher is better), and FID (lower is better). The best values in each column are highlighted in bold. Conditioning on specific modalities significantly improves performance, showcasing the advantages of guided generation.}
    \label{tab:gen_eval}
\end{table*}

}
% {\input{tables/discriminative}}

\subsection{Quantitative Results}

\paragraph{Evaluation Protocol.}
To quantitatively benchmark our results, we compare them against DiffusionSAT \cite{khanna2024diffusionsat}, the only general-purpose diffusion model for remote sensing (RS) with open training code, inference code, and weights. Since DiffusionSAT generates only optical images, we compare its performance (without any text conditioning, i.e. empty text prompt) with our model on S2L2A and S2L1C for unconditional generation.

To ensure a fair evaluation, we use a test set unseen during training. For comparison, we use Fréchet Inception Distance (FID), Precision, Recall, and F-Score.

\begin{itemize}
    \item \textbf{FID} measures the distribution distance between generated and real images by extracting deep features using a pretrained inception-v3-compat model\footnote{\url{https://github.com/toshas/torch-fidelity/releases/download/v0.2.0/weights-inception-2015-12-05-6726825d.pth}}. Lower FID indicates higher visual fidelity.
    \item \textbf{Precision} quantifies the proportion of generated images that are realistic by measuring whether each generated image has a real counterpart within its nearest neighbours feature space.
    \item \textbf{Recall} measures the diversity of generated images by checking how well they cover the distribution of real images.
    \item \textbf{F-Score} is the harmonic mean of Precision and Recall, balancing realism and diversity.
\end{itemize}

To extract image features we use a pretrained dino-v2-s-14 model\footnote{\url{https://dl.fbaipublicfiles.com/dinov2/dinov2_vits14/dinov2_vits14_pretrain.pth}}. Precision, Recall, and F-Score are computed by comparing feature distances between real and generated images, using k-nearest neighbors to determine matching quality.

We also evaluate our model's performance in conditional generation by providing one modality as input and generating the remaining channels.

\paragraph{Experimental Results.}

As shown in Table \ref{tab:gen_eval}, \ourmodel consistently outperforms DiffusionSat across all metrics on Copernicus data, demonstrating its superior ability to generate high-quality satellite images. It achieves significantly higher F-Scores, such as 36.71 for S2L1C, compared to DiffusionSat’s 1.37, highlighting the latter's limitations. Similarly, for Precision, \ourmodel reaches 71.72 for S2L2A, while DiffusionSat lags at 1.97. FID scores also confirm this gap, with \ourmodel achieving a much lower 48.49 for S2L2A versus DiffusionSat’s 216.37, indicating more realistic Sentinel-2 image generation. Conditioning further enhances \ourmodel's performance.

Interestingly, DEM is the easiest modality to generate in the unconditional case but benefits less from conditioning, suggesting weaker correlations with other modalities. In contrast, S2L1C and S2L2A exhibit a strong link, which is expected as they are different processing levels of the same satellite and originate from the same sensor, making them inherently closer \footnote{We primarily consider S2L1C and S2L2A as separate modalities to allow for greater sampling flexibility and simplify method description. However, it is clear that they are very closely related as they are derived from the same raw product.}.

While these results focus on Copernicus data, \ourmodel also generalises well to other datasets with different processing pipelines, as shown in the qualitative analysis. %\ref{sec:qualitative-results}

    {\begin{figure}[ht]
    \centering
    \includegraphics[width=\linewidth]{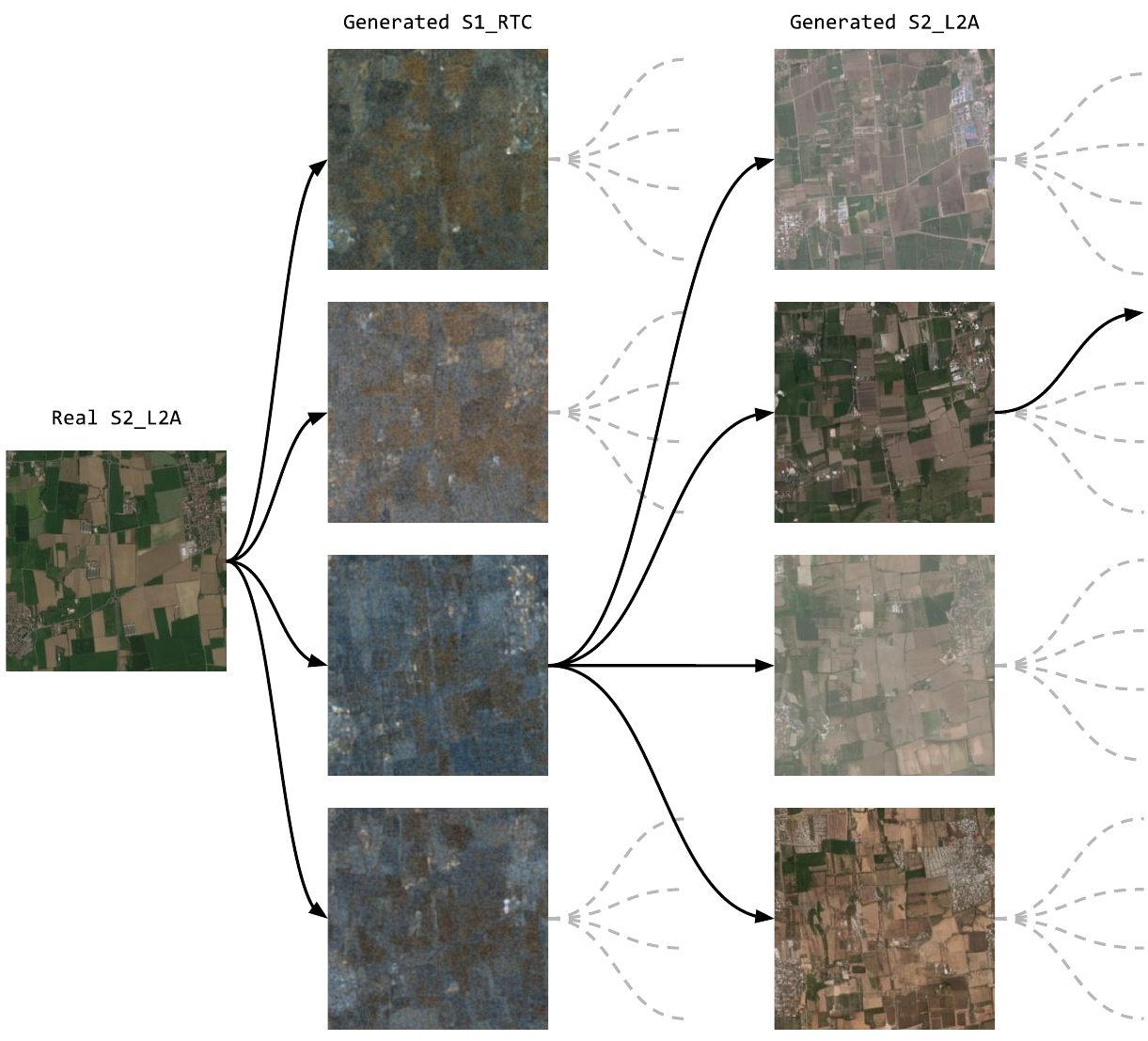}
    \caption{\textbf{Looping}. We illustrate the iterative process of conditioning the model on its own generated outputs. Starting from a real Sentinel-2 L2A image (left), the model first generates multiple corresponding Sentinel-1 RTC image (middle), which is then used to synthesize a new Sentinel-2 L2A image (right). Best viewed when zoomed in. Longer loop sequences can be found in the Supplementary Material.}
    \label{fig:short-looping}
\end{figure}}

    {\begin{figure}[ht]
    \centering
    \includegraphics[width=\linewidth]{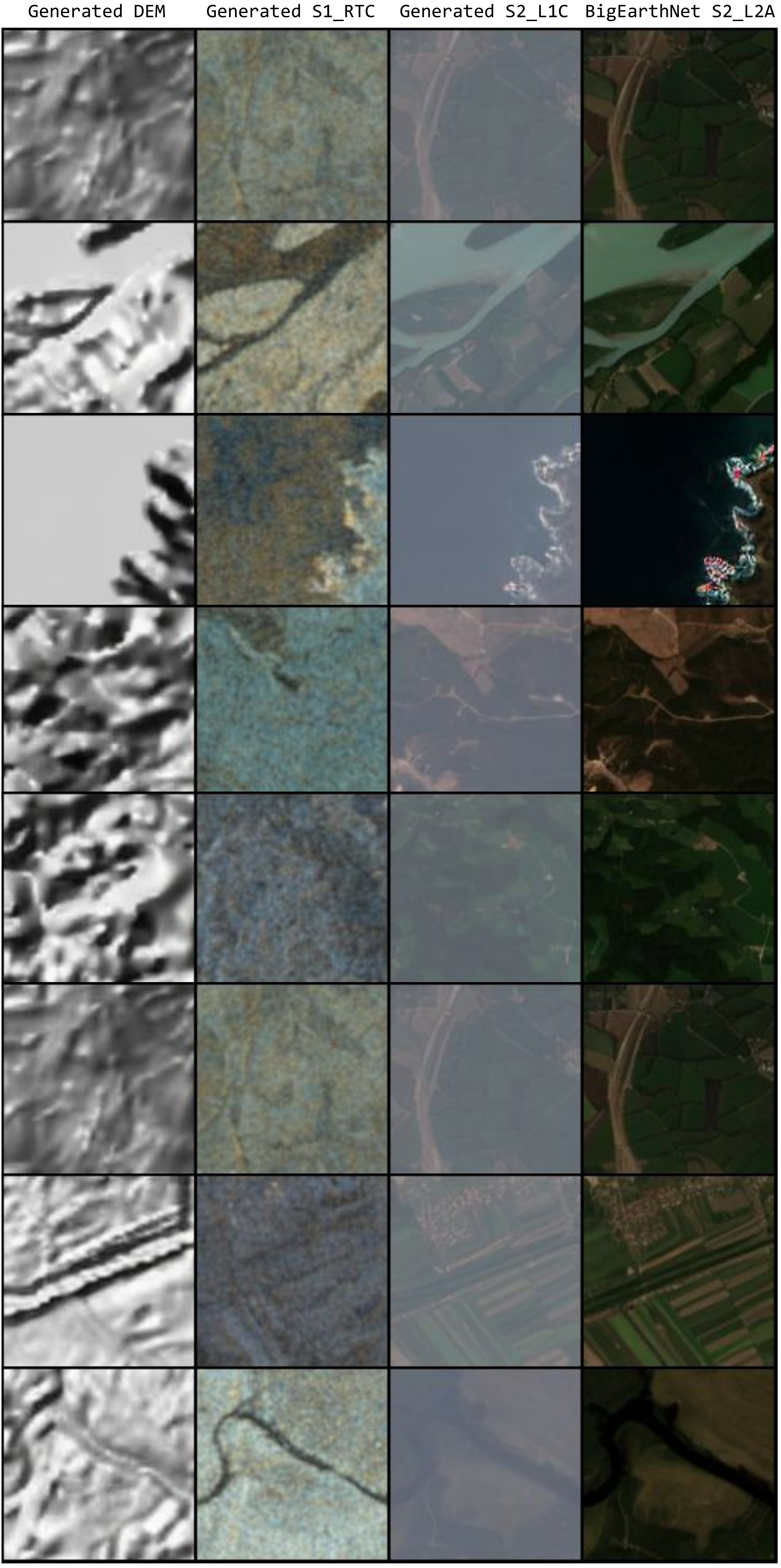}
    \caption{\ourmodel can be used to expand existing datasets containing the supported modalities, such as BigEarthNet, which has been used as a conditioning source of Sentinel-2 L2A. \ourmodel is capable of reproducing all remaining modalities despite no exposure to BigEarthNet samples in the training data.}
    \label{fig:bigearthnet}
\end{figure}}

    \subsection{Qualitative Results}
    \label{sec:qualitative-results}
   
    We qualitatively assess our model’s performance by visually inspecting its outputs under different conditioning modes. A key strength of \ourmodel is its versatility in sampling, unlocking a wide range of downstream applications. %Figure \ref{fig:all-to-all} illustrates the various conditioning combinations our model supports, offering significant flexibility.

    One application is atmospheric correction. As shown in Figure \ref{fig:atmospheric-correction}, our model can estimate atmospheric effects, effectively replicating the official L2A-level processor and enabling, for free, atmospheric effect removal. Conversely, it can synthesize atmospheric effects when conditioned on already processed S2L2A images.

    Another use case involves DEM generation. Figure \ref{fig:elevation-estimation} shows how our model generates DEMs conditioned on different data modalities. This capability is valuable for updating DEM products more frequently, providing a denser and more continuous observation of topological changes on Earth.

    Additionally, \ourmodel generates multiple plausible versions given the same input due to the stochastic nature of diffusion models, which excel at modeling data distributions. Trained on a large, global dataset, our model captures seasonal variations across different latitudes. Such emergent capabilities, illustrated in Figure \ref{fig:emergent-seasonality}, open up new possibilities, allowing to reimagine locations under conditions never observed before. Refer to the Supplementary Material for further visualisations on all the possible conditioning combinations our model supports.

    To investigate generalisation abilities we condition \ourmodel on images from the BigEarthNet dataset \cite{sumbul2019bigearthnet}. Despite differences in processing pipelines between Major TOM thumbnails and BigEarthNet, our model produces reasonable results shown in Figure \ref{fig:bigearthnet}, demonstrating its potential for expanding existing datasets to additional modalities.

    Figure \ref{fig:short-looping} shows how \ourmodel enables the creation of data generation loops by conditioning on previously synthesised data, leading to a tree of potential samples that could be mapped to every real sample - a sign of the complex relationship when mapping between highly different modalities.
    This also allows us to qualitatively observe whether the model's generative capabilities degrade as it continues to condition on its own outputs. For extended sequences of loops, we refer the reader to the Supplementary Material.
    
    In future work, we intend to quantitatively assess the extent of data degradation as the number of loop iterations increases.
    Additionally, an intriguing direction for further research is to investigate whether training the model on a mixture of generated and real data enhances its performance. This could, on its turn, reinforce the data loop, improving generalization and robustness in a self-improving manner.

\section{Conclusion}
\label{sec:conclusion}

This work introduces \ourmodel, a transformer-based diffusion model designed for multi-modal Earth observation imagery---the first to learn a joint generative distribution across four modalities. The effectiveness of the approach is demonstrated by training this prototype model on a global dataset of thumbnails of the Major TOM dataset and evaluating the quality of the generated output in a quantitative and qualitative manner.

The results indicate a high potential of this approach, particularly for the domain of Earth observation. The model can generate matched data based on any set of modalities provided as a condition and hence, it can already act as a source of priors when working with Copernicus data.

However, the main purpose of this work is to lay a foundation for a more comprehensive design, which fully supports original formatting of source data (instead of thumbnails) and can be easily adapted when a new modality is introduced through continual learning approaches. The results obtained in this study suggest that this foundation has been achieved.

\section*{Acknowledgements}
Funding for this research is supported through a Centre for Satellite Data in Environmental Science (SENSE) CDT studentship (NE/T00939X/1). This work used JASMIN, the
UK’s collaborative data analysis environment~\url{https://jasmin.ac.uk}~\cite{jasmin}.

\section{Appendix}
\label{sec:appendix_section}

In this section, we provide additional qualitative results and visualizations to further illustrate the capabilities of \ourmodel.

Figure \ref{fig:unconditional} presents examples of unconditional synthesis, demonstrating the diversity and variability captured by \ourmodel across different modalities. These results highlight the model’s ability to capture the full spectrum of the training data distribution.

{% Use figure* for multi-column figure
\begin{figure*}[ht]
    \centering
    \includegraphics[width=\linewidth]{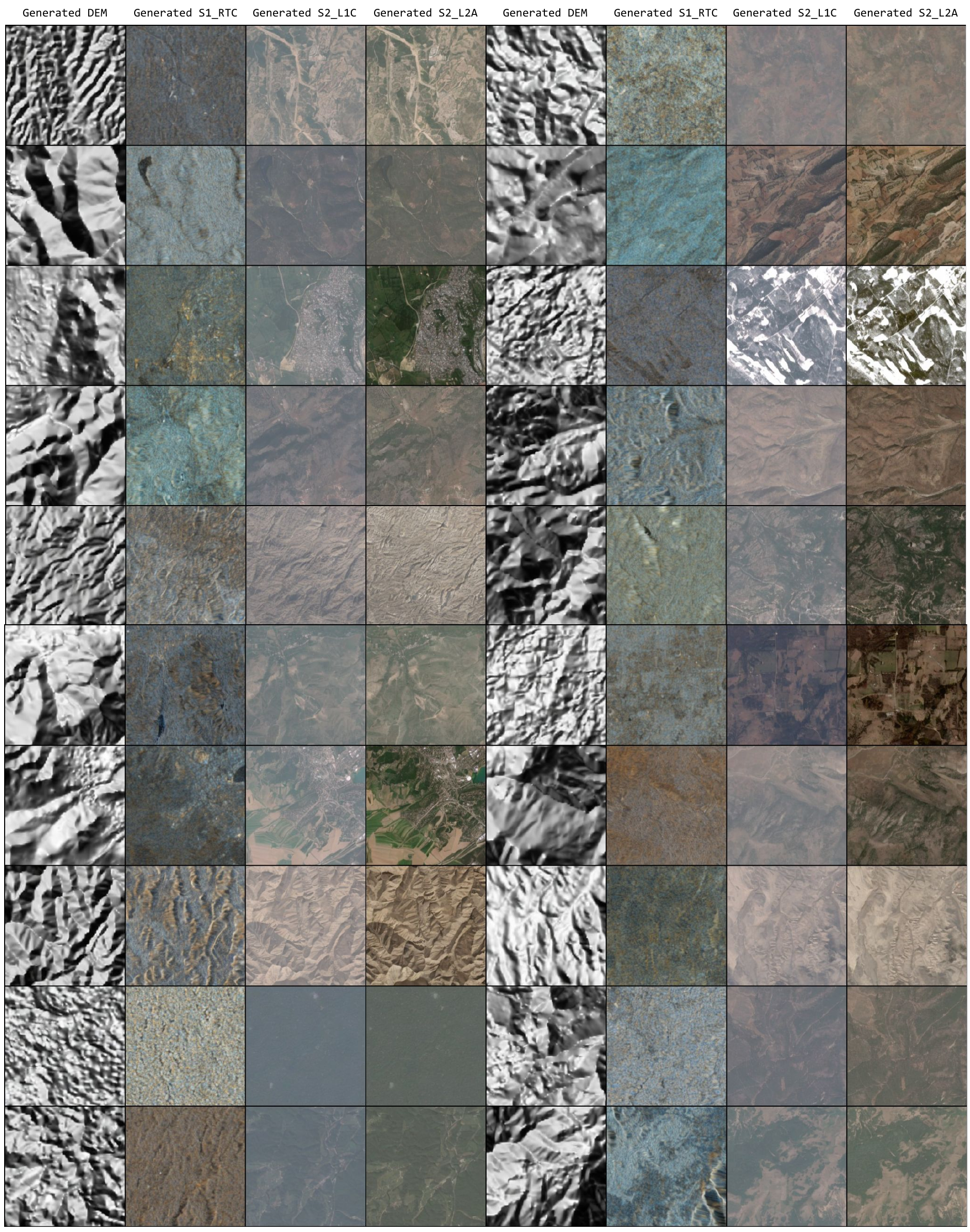}
    \caption{Unconditional joint generation of all modalities. Rows represent independent samplings (in groups of 4). The model generations exhibit high diversity, consistent with the rich and global data used for training.}
    \label{fig:unconditional}
\end{figure*}
}

Figure \ref{fig:long-looping} showcases an extended loop between the S2L2A and S1RTC modalities. This visualization provides insight into how the model maintains consistency across generative steps. Future work will focus on a quantitative assessment of image quality, particularly analyzing potential degradation over iterative generations.

{% Use figure* for multi-column figure
\begin{figure*}[ht]
    \centering
    \includegraphics[width=\linewidth]{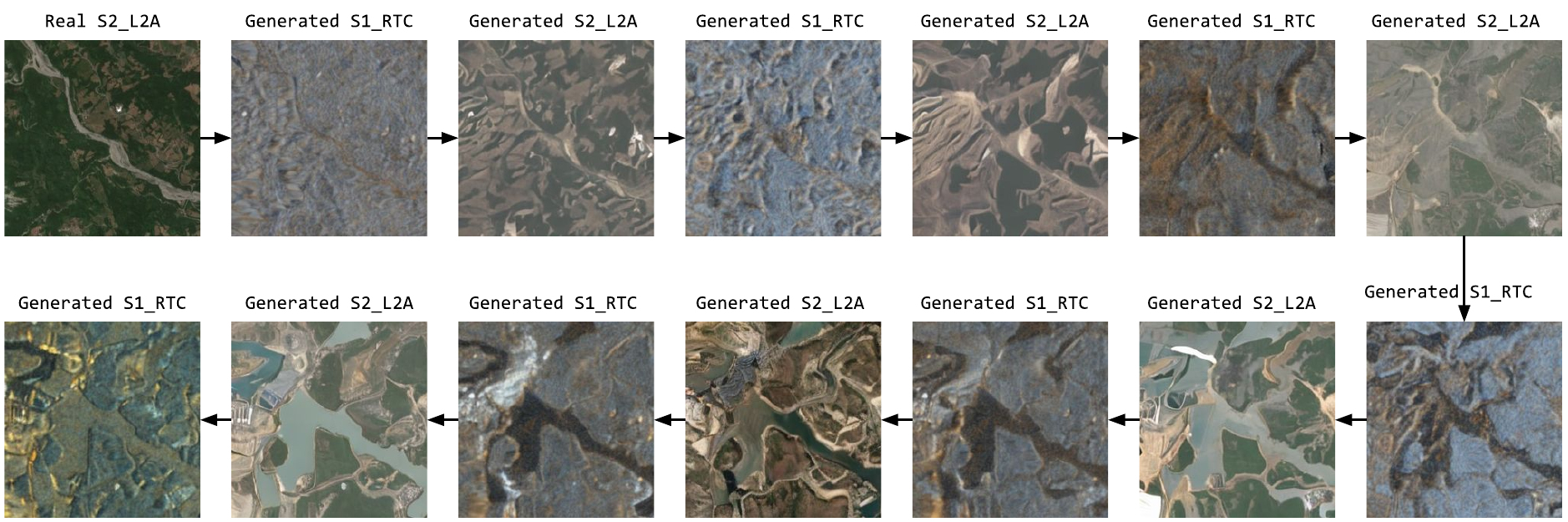}
    \caption{To analyse the model robustness to generation degradation, we perform a generation chain, starting from a real S2L2A image and iteratively conditioning the model on the previously generated image. For illustrative purposes, we only show a loop between S2L2A and S1RTC modalities repeatedly.
    }
    \label{fig:long-looping}
\end{figure*}
}

For completeness, Figure \ref{fig:all-to-all} illustrates all possible conditioning combinations for \ourmodel. Each row represents an independent example, where the red boxes indicate the input modalities used to condition the generation of the remaining ones. The left side of the figure provides real images (both conditioned and unconditioned) for reference, while the right side displays the generated outputs. This visualization highlights the flexibility and robustness of \ourmodel in handling diverse cross-modal synthesis tasks, effectively leveraging many different input configurations to produce high-quality outputs.

{% Use figure* for multi-column figure
\begin{figure*}[ht]
    \centering
    \includegraphics[width=0.7\linewidth]{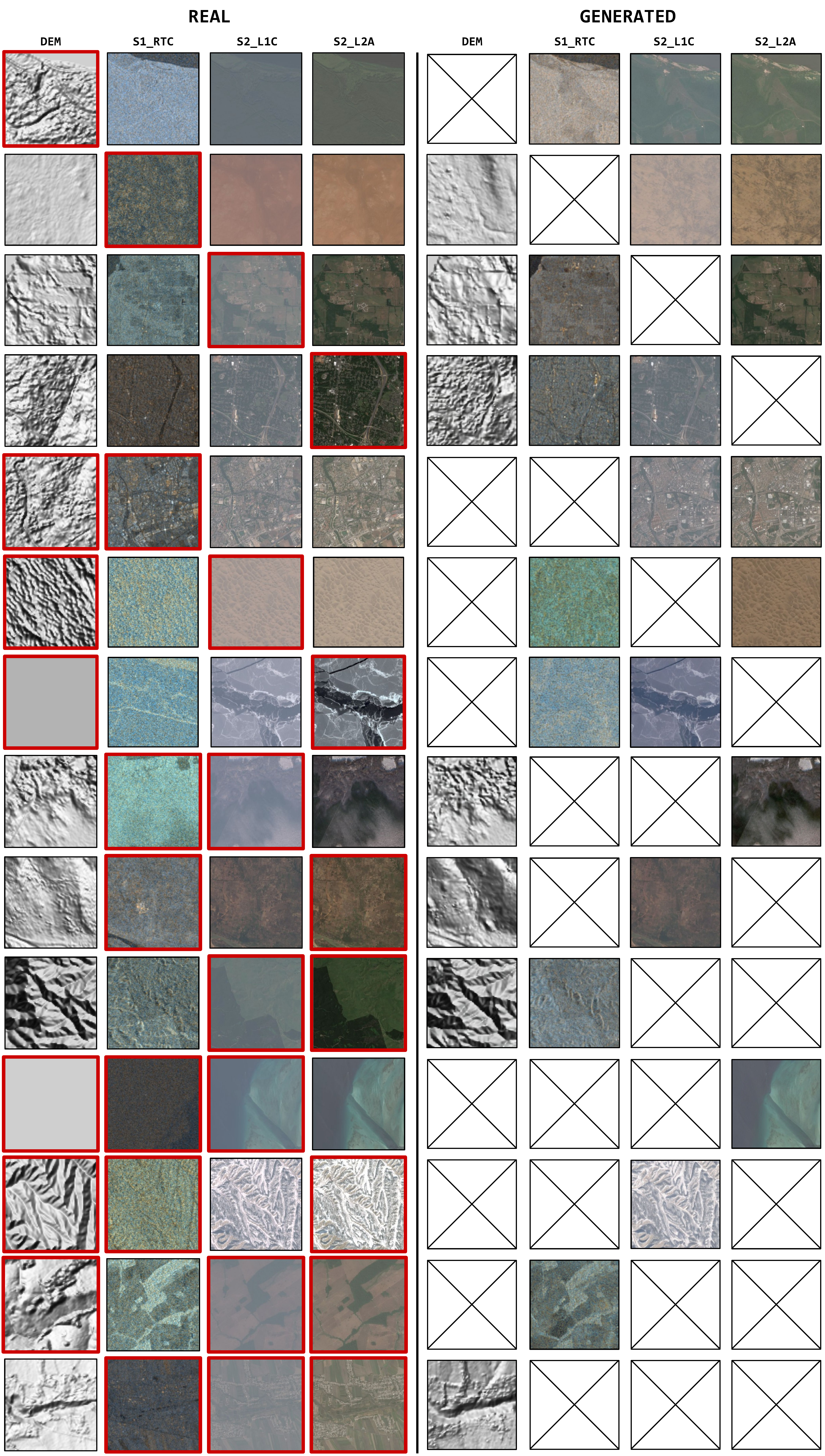}
    \caption{Illustration of all the possible combinations for conditioning \ourmodel. We show all the possible ways of conditioning our model. Each row represents an independent example. The red boxes (\textcolor{myred}{$\blacksquare$}) indicate the modalities used as input to condition the generation of the remaining modalities. The right side of the figure displays the generated outputs, while the left side includes the corresponding real images (both conditioned and unconditioned) for comparison.}
    \label{fig:all-to-all}
\end{figure*}
}

% \textcolor{red}{We can probably remove the figures of the dataset coverage for train and test split. Honestly, I don't think they add much.}

% {\input{figs/dataset-coverage}}


{\small
\begin{thebibliography}{57}
\providecommand{\natexlab}[1]{#1}
\providecommand{\url}[1]{\texttt{#1}}
\expandafter\ifx\csname urlstyle\endcsname\relax
  \providecommand{\doi}[1]{doi: #1}\else
  \providecommand{\doi}{doi: \begingroup \urlstyle{rm}\Url}\fi

\bibitem[Agency(2023)]{copernicus2022}
European~Space Agency.
\newblock Copernicus sentinel data access annual report 2022, 2023.
\newblock Accessed: 2025-03-06.

\bibitem[Apicella et~al.(2022)Apicella, De~Martino, and Quarati]{apicella2022copernicus}
Lorenza Apicella, Monica De~Martino, and Alfonso Quarati.
\newblock Copernicus user uptake: From data to applications.
\newblock \emph{ISPRS International Journal of Geo-Information}, 11\penalty0 (2):\penalty0 121, 2022.

\bibitem[Aschbacher(2017)]{aschbacher2017esa}
Josef Aschbacher.
\newblock Esa’s earth observation strategy and copernicus.
\newblock In \emph{Satellite earth observations and their impact on society and policy}, pages 81--86. Springer Singapore Singapore, 2017.

\bibitem[Bao et~al.(2023{\natexlab{a}})Bao, Nie, Xue, Cao, Li, Su, and Zhu]{Bao_2023_CVPR}
Fan Bao, Shen Nie, Kaiwen Xue, Yue Cao, Chongxuan Li, Hang Su, and Jun Zhu.
\newblock All are worth words: A vit backbone for diffusion models.
\newblock In \emph{Proceedings of the IEEE/CVF Conference on Computer Vision and Pattern Recognition (CVPR)}, pages 22669--22679, 2023{\natexlab{a}}.

\bibitem[Bao et~al.(2023{\natexlab{b}})Bao, Nie, Xue, Li, Pu, Wang, Yue, Cao, Su, and Zhu]{pmlr-v202-bao23a}
Fan Bao, Shen Nie, Kaiwen Xue, Chongxuan Li, Shi Pu, Yaole Wang, Gang Yue, Yue Cao, Hang Su, and Jun Zhu.
\newblock One transformer fits all distributions in multi-modal diffusion at scale.
\newblock In \emph{Proceedings of the 40th International Conference on Machine Learning}, pages 1692--1717. PMLR, 2023{\natexlab{b}}.

\bibitem[Borne-Pons et~al.(2025)Borne-Pons, Czerkawski, Martin, and Rouffet]{bornepons2025mesa}
Paul Borne-Pons, Mikolaj Czerkawski, Rosalie Martin, and Romain Rouffet.
\newblock {MESA}: Text-driven terrain generation using latent diffusion and global copernicus data.
\newblock In \emph{CVPRW 2025 MORSE Workshop}, 2025.

\bibitem[Czerkawski and Francis(2023)]{czerkawski2023laion}
Mikolaj Czerkawski and Alistair Francis.
\newblock From laion-5b to laion-eo: Filtering billions of images using anchor datasets for satellite image extraction.
\newblock In \emph{"Towards the Next Generation of Computer Vision Datasets: DataComp Track" Workshop at the IEEE/CVF International Conference on Computer Vision (ICCV)}, 2023.

\bibitem[Czerkawski and Tachtatzis(2024)]{Czerkawski_2024}
Mikolaj Czerkawski and Christos Tachtatzis.
\newblock Exploring the capability of text-to-image diffusion models with structural edge guidance for multispectral satellite image inpainting.
\newblock \emph{IEEE Geoscience and Remote Sensing Letters}, 21:\penalty0 1–5, 2024.

\bibitem[Dong et~al.(2024)Dong, Yuan, Luo, Chen, Zhang, Zhang, Li, Zheng, and Fu]{dong2024building}
Runmin Dong, Shuai Yuan, Bin Luo, Mengxuan Chen, Jinxiao Zhang, Lixian Zhang, Weijia Li, Juepeng Zheng, and Haohuan Fu.
\newblock Building bridges across spatial and temporal resolutions: Reference-based super-resolution via change priors and conditional diffusion model.
\newblock In \emph{Proceedings of the IEEE/CVF Conference on Computer Vision and Pattern Recognition}, pages 27684--27694, 2024.

\bibitem[Enescu and Sahbi(2024)]{enescu2024conditional}
Victor Enescu and Hichem Sahbi.
\newblock Conditional normalizing flows for nonlinear remote sensing image augmentation and classification.
\newblock In \emph{IGARSS 2024-2024 IEEE International Geoscience and Remote Sensing Symposium}, pages 10264--10268. IEEE, 2024.

\bibitem[Espinosa and Crowley(2023)]{espinosa_2023_8_mapsat}
Miguel Espinosa and Elliot~J. Crowley.
\newblock Generate your own scotland: Satellite image generation conditioned on maps.
\newblock \emph{NeurIPS 2023 Workshop on Diffusion Models}, 2023.

\bibitem[et~al.(2021)]{Bommasani2021FoundationModels}
Rishi~Bommasani et al.
\newblock On the opportunities and risks of foundation models.
\newblock \emph{ArXiv}, 2021.

\bibitem[Fooladgar and Kasaei(2019)]{fooladgar2019multi}
Fahimeh Fooladgar and Shohreh Kasaei.
\newblock Multi-modal attention-based fusion model for semantic segmentation of rgb-depth images.
\newblock \emph{arXiv preprint arXiv:1912.11691}, 2019.

\bibitem[Francis and Czerkawski(2024)]{francis2024major}
Alistair Francis and Mikolaj Czerkawski.
\newblock Major tom: Expandable datasets for earth observation.
\newblock In \emph{IGARSS 2024-2024 IEEE International Geoscience and Remote Sensing Symposium}, pages 2935--2940. IEEE, 2024.

\bibitem[Fuller et~al.(2023)Fuller, Millard, and Green]{fuller2023croma}
Anthony Fuller, Koreen Millard, and James~R. Green.
\newblock Croma: Remote sensing representations with contrastive radar-optical masked autoencoders, 2023.

\bibitem[Gao et~al.(2023)Gao, Lin, Zhou, Liu, Xie, Ge, and Jiang]{gao2023textpainter}
Yifan Gao, Jinpeng Lin, Min Zhou, Chuanbin Liu, Hongtao Xie, Tiezheng Ge, and Yuning Jiang.
\newblock Textpainter: Multimodal text image generation with visual-harmony and text-comprehension for poster design.
\newblock In \emph{Proceedings of the 31st ACM International Conference on Multimedia}, pages 7236--7246, 2023.

\bibitem[Han et~al.(2024)Han, Cao, Poon, and Navigli]{han2024multimodal}
Soyeon~Caren Han, Feiqi Cao, Josiah Poon, and Roberto Navigli.
\newblock Multimodal large language models and tunings: Vision, language, sensors, audio, and beyond.
\newblock In \emph{Proceedings of the 32nd ACM International Conference on Multimedia}, pages 11294--11295, 2024.

\bibitem[He et~al.(2023)He, Li, Yuan, et~al.]{he2023tdiffde}
Jiang He, Yajie Li, Qiangqiang Yuan, et~al.
\newblock Tdiffde: A truncated diffusion model for remote sensing hyperspectral image denoising.
\newblock \emph{arXiv preprint arXiv:2311.13622}, 2023.

\bibitem[Ho et~al.(2020)Ho, Jain, and Abbeel]{ho2020denoising}
Jonathan Ho, Ajay Jain, and Pieter Abbeel.
\newblock Denoising diffusion probabilistic models.
\newblock \emph{Advances in neural information processing systems}, 33:\penalty0 6840--6851, 2020.

\bibitem[Hong et~al.(2024)Hong, Zhang, Li, Li, Li, Yao, Yokoya, Li, Ghamisi, Jia, Plaza, Gamba, Benediktsson, and Chanussot]{Hong2024spectralgpt}
Danfeng Hong, Bing Zhang, Xuyang Li, Yuxuan Li, Chenyu Li, Jing Yao, Naoto Yokoya, Hao Li, Pedram Ghamisi, Xiuping Jia, Antonio Plaza, Paolo Gamba, Jon~Atli Benediktsson, and Jocelyn Chanussot.
\newblock Spectralgpt: Spectral remote sensing foundation model.
\newblock \emph{IEEE Transactions on Pattern Analysis and Machine Intelligence}, 2024.

\bibitem[Hu et~al.(2024)Hu, Chan, Su, Chen, Li, Sohn, Zhao, Ben, Gong, Cohen, Chang, and Jia]{Hu_2024_CVPR}
Hexiang Hu, Kelvin~C.K. Chan, Yu-Chuan Su, Wenhu Chen, Yandong Li, Kihyuk Sohn, Yang Zhao, Xue Ben, Boqing Gong, William Cohen, Ming-Wei Chang, and Xuhui Jia.
\newblock Instruct-imagen: Image generation with multi-modal instruction.
\newblock In \emph{Proceedings of the IEEE/CVF Conference on Computer Vision and Pattern Recognition (CVPR)}, pages 4754--4763, 2024.

\bibitem[Huang et~al.(2023)Huang, Chan, Jiang, and Liu]{Huang_2023_CVPR}
Ziqi Huang, Kelvin~C.K. Chan, Yuming Jiang, and Ziwei Liu.
\newblock Collaborative diffusion for multi-modal face generation and editing.
\newblock In \emph{Proceedings of the IEEE/CVF Conference on Computer Vision and Pattern Recognition (CVPR)}, pages 6080--6090, 2023.

\bibitem[Jiang et~al.(2019)Jiang, Wang, Yi, Wang, Lu, and Jiang]{jiang2019edge}
Kui Jiang, Zhongyuan Wang, Peng Yi, Guangcheng Wang, Tao Lu, and Junjun Jiang.
\newblock Edge-enhanced gan for remote sensing image superresolution.
\newblock \emph{IEEE Transactions on Geoscience and Remote Sensing}, 57\penalty0 (8):\penalty0 5799--5812, 2019.

\bibitem[Khanna et~al.(2024)Khanna, Liu, Zhou, Meng, Rombach, Burke, Lobell, and Ermon]{khanna2024diffusionsat}
Samar Khanna, Patrick Liu, Linqi Zhou, Chenlin Meng, Robin Rombach, Marshall Burke, David~B. Lobell, and Stefano Ermon.
\newblock Diffusionsat: A generative foundation model for satellite imagery.
\newblock In \emph{The Twelfth International Conference on Learning Representations}, 2024.

\bibitem[Kumar et~al.(2023)Kumar, Azamathulla, Sharma, Mehta, and Maharaj]{kumar2023state}
Vijendra Kumar, Hazi~Md Azamathulla, Kul~Vaibhav Sharma, Darshan~J Mehta, and Kiran~Tota Maharaj.
\newblock The state of the art in deep learning applications, challenges, and future prospects: A comprehensive review of flood forecasting and management.
\newblock \emph{Sustainability}, 15\penalty0 (13):\penalty0 10543, 2023.

\bibitem[Lawrence et~al.(2013)Lawrence, Bennett, Churchill, Juckes, Kershaw, Pascoe, Pepler, Pritchard, and Stephens]{jasmin}
Bryan~N. Lawrence, Victoria~L. Bennett, James Churchill, Martin Juckes, Philip Kershaw, Stephen Pascoe, Sam Pepler, Matthew Pritchard, and Ag Stephens.
\newblock Storing and manipulating environmental big data with jasmin.
\newblock In \emph{IEEE Big Data}, pages 1--5, San Francisco, 2013. IEEE.

\bibitem[Le~Bellier and Audebert(2024)]{le2024detecting}
Georges Le~Bellier and Nicolas Audebert.
\newblock Detecting out-of-distribution earth observation images with diffusion models.
\newblock In \emph{Proceedings of the IEEE/CVF Conference on Computer Vision and Pattern Recognition}, pages 481--491, 2024.

\bibitem[Li et~al.(2022)Li, Hong, Gao, Yao, Zheng, Zhang, and Chanussot]{li2022deep}
Jiaxin Li, Danfeng Hong, Lianru Gao, Jing Yao, Ke Zheng, Bing Zhang, and Jocelyn Chanussot.
\newblock Deep learning in multimodal remote sensing data fusion: A comprehensive review.
\newblock \emph{International Journal of Applied Earth Observation and Geoinformation}, 112:\penalty0 102926, 2022.

\bibitem[Li et~al.(2021)Li, Du, Huang, and Tan]{li2021deep}
Xinghua Li, Zhengshun Du, Yanyuan Huang, and Zhenyu Tan.
\newblock A deep translation (gan) based change detection network for optical and sar remote sensing images.
\newblock \emph{ISPRS Journal of Photogrammetry and Remote Sensing}, 179:\penalty0 14--34, 2021.

\bibitem[Li et~al.(2023)Li, B{\"u}rgi, Ma, Noh, Wald, and Xu]{li2023disasternet}
Xuechun Li, Paula~M B{\"u}rgi, Wei Ma, Hae~Young Noh, David~Jay Wald, and Susu Xu.
\newblock Disasternet: Causal bayesian networks with normalizing flows for cascading hazards estimation from satellite imagery.
\newblock In \emph{Proceedings of the 29th ACM SIGKDD Conference on Knowledge Discovery and Data Mining}, pages 4391--4403, 2023.

\bibitem[Liu et~al.(2024)Liu, Yue, Xia, Ghamisi, Xie, and Fang]{liu2024diffusion}
Yidan Liu, Jun Yue, Shaobo Xia, Pedram Ghamisi, Weiying Xie, and Leyuan Fang.
\newblock Diffusion models meet remote sensing: Principles, methods, and perspectives.
\newblock \emph{arXiv preprint arXiv:2404.08926}, 2024.

\bibitem[Marsocci et~al.(2024)Marsocci, Jia, Bellier, Kerekes, Zeng, Hafner, Gerard, Brune, Yadav, Shibli, Fang, Ban, Vergauwen, Audebert, and Nascetti]{marsocci2024pangaea}
Valerio Marsocci, Yuru Jia, Georges~Le Bellier, David Kerekes, Liang Zeng, Sebastian Hafner, Sebastian Gerard, Eric Brune, Ritu Yadav, Ali Shibli, Heng Fang, Yifang Ban, Maarten Vergauwen, Nicolas Audebert, and Andrea Nascetti.
\newblock Pangaea: A global and inclusive benchmark for geospatial foundation models, 2024.

\bibitem[Miller et~al.(2024)Miller, Pelletier, and Webb]{miller2024deep}
Lynn Miller, Charlotte Pelletier, and Geoffrey~I Webb.
\newblock Deep learning for satellite image time-series analysis: A review.
\newblock \emph{IEEE Geoscience and Remote Sensing Magazine}, 2024.

\bibitem[Nedungadi et~al.(2024)Nedungadi, Kariryaa, Oehmcke, Belongie, Igel, and Lang]{nedungadi2024mmearth}
Vishal Nedungadi, Ankit Kariryaa, Stefan Oehmcke, Serge Belongie, Christian Igel, and Nico Lang.
\newblock Mmearth: Exploring multi-modal pretext tasks for geospatial representation learning.
\newblock \emph{arXiv preprint arXiv:2405.02771}, 2024.

\bibitem[Pang et~al.(2024)Pang, Rui, Cui, Wang, Meng, and Cao]{pang2024hir}
Li Pang, Xiangyu Rui, Long Cui, Hongzhong Wang, Deyu Meng, and Xiangyong Cao.
\newblock Hir-diff: Unsupervised hyperspectral image restoration via improved diffusion models.
\newblock In \emph{Proceedings of the IEEE/CVF Conference on Computer Vision and Pattern Recognition}, pages 3005--3014, 2024.

\bibitem[Radford et~al.(2021)Radford, Kim, Hallacy, Ramesh, Goh, Agarwal, Sastry, Askell, Mishkin, Clark, et~al.]{radford2021learning}
Alec Radford, Jong~Wook Kim, Chris Hallacy, Aditya Ramesh, Gabriel Goh, Sandhini Agarwal, Girish Sastry, Amanda Askell, Pamela Mishkin, Jack Clark, et~al.
\newblock Learning transferable visual models from natural language supervision.
\newblock In \emph{International conference on machine learning}, pages 8748--8763, 2021.

\bibitem[Ren et~al.(2024)Ren, Yao, Li, Sun, and Hou]{ren2024timechat}
Shuhuai Ren, Linli Yao, Shicheng Li, Xu Sun, and Lu Hou.
\newblock Timechat: A time-sensitive multimodal large language model for long video understanding.
\newblock In \emph{Proceedings of the IEEE/CVF Conference on Computer Vision and Pattern Recognition}, pages 14313--14323, 2024.

\bibitem[Rombach et~al.(2022)Rombach, Blattmann, Lorenz, Esser, and Ommer]{Rombach_2022_CVPR}
Robin Rombach, Andreas Blattmann, Dominik Lorenz, Patrick Esser, and Bj\"orn Ommer.
\newblock High-resolution image synthesis with latent diffusion models.
\newblock In \emph{Proceedings of the IEEE/CVF Conference on Computer Vision and Pattern Recognition (CVPR)}, 2022.

\bibitem[Ruan et~al.(2023)Ruan, Ma, Yang, He, Liu, Fu, Yuan, Jin, and Guo]{Ruan_2023_CVPR}
Ludan Ruan, Yiyang Ma, Huan Yang, Huiguo He, Bei Liu, Jianlong Fu, Nicholas~Jing Yuan, Qin Jin, and Baining Guo.
\newblock Mm-diffusion: Learning multi-modal diffusion models for joint audio and video generation.
\newblock In \emph{Proceedings of the IEEE/CVF Conference on Computer Vision and Pattern Recognition (CVPR)}, pages 10219--10228, 2023.

\bibitem[Ruiz et~al.(2023)Ruiz, Li, Jampani, Pritch, Rubinstein, and Aberman]{Ruiz_2023_CVPR}
Nataniel Ruiz, Yuanzhen Li, Varun Jampani, Yael Pritch, Michael Rubinstein, and Kfir Aberman.
\newblock Dreambooth: Fine tuning text-to-image diffusion models for subject-driven generation.
\newblock In \emph{Proceedings of the IEEE/CVF Conference on Computer Vision and Pattern Recognition (CVPR)}, pages 22500--22510, 2023.

\bibitem[Sanguigni et~al.(2023)Sanguigni, Czerkawski, Papa, Amerini, and Saux]{Sanguini2023}
Fulvio Sanguigni, Mikolaj Czerkawski, Lorenzo Papa, Irene Amerini, and Bertrand~Le Saux.
\newblock Diffusion models for earth observation use-cases: from cloud removal to urban change detection.
\newblock In \emph{Proceedings of the 2023 conference on Big Data from Space (BiDS’23): from foresight to impact : 6 9 November 2023, Austrian Center, Vienna.} European Commission. Joint Research Centre., 2023.

\bibitem[Sumbul et~al.(2019)Sumbul, Charfuelan, Demir, and Markl]{sumbul2019bigearthnet}
Gencer Sumbul, Marcela Charfuelan, Beg{\"u}m Demir, and Volker Markl.
\newblock Bigearthnet: A large-scale benchmark archive for remote sensing image understanding.
\newblock In \emph{IGARSS 2019-2019 IEEE international geoscience and remote sensing symposium}, pages 5901--5904. IEEE, 2019.

\bibitem[Tang et~al.(2024)Tang, Cao, Hou, Jiang, Liu, and Meng]{tang2024crs}
Datao Tang, Xiangyong Cao, Xingsong Hou, Zhongyuan Jiang, Junmin Liu, and Deyu Meng.
\newblock Crs-diff: Controllable remote sensing image generation with diffusion model.
\newblock \emph{IEEE Transactions on Geoscience and Remote Sensing}, 2024.

\bibitem[Van~Etten et~al.(2018)Van~Etten, Lindenbaum, and Bacastow]{van2018spacenet}
Adam Van~Etten, Dave Lindenbaum, and Todd~M Bacastow.
\newblock Spacenet: A remote sensing dataset and challenge series.
\newblock \emph{arXiv preprint arXiv:1807.01232}, 2018.

\bibitem[Wang and Sun(2025)]{wang2025semantic}
Ce Wang and Wanjie Sun.
\newblock Semantic guided large scale factor remote sensing image super-resolution with generative diffusion prior.
\newblock \emph{ISPRS Journal of Photogrammetry and Remote Sensing}, 220:\penalty0 125--138, 2025.

\bibitem[Wang et~al.(2024)Wang, Song, Wei, Xian, Shi, and Lin]{wang2024idf}
Meilin Wang, Yexing Song, Pengxu Wei, Xiaoyu Xian, Yukai Shi, and Liang Lin.
\newblock Idf-cr: Iterative diffusion process for divide-and-conquer cloud removal in remote-sensing images.
\newblock \emph{IEEE Transactions on Geoscience and Remote Sensing}, 2024.

\bibitem[Wang et~al.(2023)Wang, Sun, Chehri, and Song]{wang2023review}
Xuan Wang, Lijun Sun, Abdellah Chehri, and Yongchao Song.
\newblock A review of gan-based super-resolution reconstruction for optical remote sensing images.
\newblock \emph{Remote Sensing}, 15\penalty0 (20):\penalty0 5062, 2023.

\bibitem[Wu et~al.(2023)Wu, Ni, Wang, and Zhang]{wu2023conditional}
Hanlin Wu, Ning Ni, Shan Wang, and Libao Zhang.
\newblock Conditional stochastic normalizing flows for blind super-resolution of remote sensing images.
\newblock \emph{IEEE Transactions on Geoscience and Remote Sensing}, 61:\penalty0 1--16, 2023.

\bibitem[Wu et~al.(2025)Wu, Zhong, Xing, Lai, Liu, Chen, Wang, Zhu, Lu, Lu, et~al.]{wu2025visionllm}
Jiannan Wu, Muyan Zhong, Sen Xing, Zeqiang Lai, Zhaoyang Liu, Zhe Chen, Wenhai Wang, Xizhou Zhu, Lewei Lu, Tong Lu, et~al.
\newblock Visionllm v2: An end-to-end generalist multimodal large language model for hundreds of vision-language tasks.
\newblock \emph{Advances in Neural Information Processing Systems}, 37:\penalty0 69925--69975, 2025.

\bibitem[Yang et~al.(2024)Yang, Zhang, Wang, and Xie]{yang2024mma}
Lingxiao Yang, Ru-Yuan Zhang, Yanchen Wang, and Xiaohua Xie.
\newblock Mma: Multi-modal adapter for vision-language models.
\newblock In \emph{Proceedings of the IEEE/CVF Conference on Computer Vision and Pattern Recognition}, pages 23826--23837, 2024.

\bibitem[Ye et~al.(2024)Ye, Yu, Shao, Xie, Torr, and Cao]{ye2024cat}
Qilang Ye, Zitong Yu, Rui Shao, Xinyu Xie, Philip Torr, and Xiaochun Cao.
\newblock Cat: Enhancing multimodal large language model to answer questions in dynamic audio-visual scenarios.
\newblock In \emph{European Conference on Computer Vision}, pages 146--164. Springer, 2024.

\bibitem[Yu et~al.(2025)Yu, Liu, Liu, Shi, and Zou]{metaearth}
Zhiping Yu, Chenyang Liu, Liqin Liu, Zhenwei Shi, and Zhengxia Zou.
\newblock Metaearth: A generative foundation model for global-scale remote sensing image generation.
\newblock \emph{IEEE Transactions on Pattern Analysis and Machine Intelligence}, 47\penalty0 (3):\penalty0 1764--1781, 2025.

\bibitem[Zhang(2010)]{zhang2010multi}
Jixian Zhang.
\newblock Multi-source remote sensing data fusion: status and trends.
\newblock \emph{International journal of image and data fusion}, 1\penalty0 (1):\penalty0 5--24, 2010.

\bibitem[Zhang et~al.(2024)Zhang, Huang, Jin, and Lu]{zhang2024vision}
Jingyi Zhang, Jiaxing Huang, Sheng Jin, and Shijian Lu.
\newblock Vision-language models for vision tasks: A survey.
\newblock \emph{IEEE Transactions on Pattern Analysis and Machine Intelligence}, 2024.

\bibitem[Zhao et~al.(2021)Zhao, Gong, Fu, and Tao]{zhao2021adaptive}
Shanshan Zhao, Mingming Gong, Huan Fu, and Dacheng Tao.
\newblock Adaptive context-aware multi-modal network for depth completion.
\newblock \emph{IEEE Transactions on Image Processing}, 30:\penalty0 5264--5276, 2021.

\bibitem[Zheng et~al.(2024)Zheng, Ermon, Kim, Zhang, and Zhong]{zheng2024changen2}
Zhuo Zheng, Stefano Ermon, Dongjun Kim, Liangpei Zhang, and Yanfei Zhong.
\newblock Changen2: Multi-temporal remote sensing generative change foundation model.
\newblock \emph{IEEE Transactions on Pattern Analysis and Machine Intelligence}, 2024.

\bibitem[Zou et~al.(2024)Zou, Li, Xing, Zhang, Wang, Jin, and Tao]{zou2024diffcr}
Xuechao Zou, Kai Li, Junliang Xing, Yu Zhang, Shiying Wang, Lei Jin, and Pin Tao.
\newblock Diffcr: A fast conditional diffusion framework for cloud removal from optical satellite images.
\newblock \emph{IEEE Transactions on Geoscience and Remote Sensing}, 62:\penalty0 1--14, 2024.

\end{thebibliography}

}
\end{document}